# A Method for Ethical AI in Defence: A case study on developing trustworthy autonomous systems

Tara Roberson[1]*, Stephen Bornstein[2], Rain Liivoja[3], Simon Ng[1], Jason Scholz[1], Kate Devitt[1]
*corresponding author, tara.roberson@tasdcrc.com.au

[1]Trusted Autonomous Systems, Defence Cooperative Research Centre, Brisbane, Australia
[2]Athena AI, Brisbane, Australia
[3]The University of Queensland Law School, Brisbane, Australia

**Abstract**
What does it mean to be responsible and responsive when developing and deploying trusted autonomous systems in Defence? In this short reflective article, we describe a case study of building a trusted autonomous system – Athena AI - within an industry-led, government-funded project with diverse collaborators and stakeholders. Using this case study, we draw out lessons on the value and impact of embedding responsible research and innovation-aligned, ethics-by-design approaches and principles throughout the development of technology at high translation readiness levels.

*Introduction*

Smart autonomous robotic systems that provide the asymmetric advantage to fight and win have captured the imagination of Defence. Human-autonomy teams have wide-ranging Defence applications including warfighting, logistics, humanitarian support, and the completion of dangerous jobs. Australia's Chief Defence Scientist has said that these technologies offer many benefits while emphasising that "consideration of ethical aspects needs to occur in parallel with technology development" (Ziesing, 2021).

But what does it mean to be responsible and responsive when developing and deploying trusted autonomous systems in Defence? This short article reflects on our experience of applying a method for achieving responsible research and innovation (RRI) outcomes within an industry-led, government-funded project with diverse collaborators and stakeholders.

We describe the case study focused on the design and development of a trusted autonomous system – Athena AI – which aims to augment human ethical and legal decision-making on the battlefield. Athena AI uses AI to quickly and accurately identify civilians and other protected objects to reduce the 'fog of war' and improve abidance with international humanitarian law. Using this case study, we draw out lessons on the value and impact of embedding RRI-aligned ethics-by-design approaches and principles throughout the development of technology at high translation readiness levels.

*RRI in industry and Defence*

RRI research from academia centred around trusted autonomous systems has considered societal and technical challenges posed by these emerging technologies. This includes investigating avenues for the development of ethical and safe human-AI partnerships (Ramchurn, Stein, and Jennings, 2021), identifying tools to understand responsibility for humans and AI agents

(Yazdanpanah et al., 2021), and surveying explainability methods for autonomous systems (Omeiza et al., 2021). When it comes to implementing RRI within industry settings, we face specific obstacles (Martinuzzi, Blok, Brem, Stahl, & Schönherr, 2018). For one, timescales for industry-based innovation actors are much shorter than those of researchers within academia (Tait, 2017). Potential reliance on research funding decisions creates an additional challenge. In short, engaging in RRI from an industry standpoint can be expensive in both time and money, which can hamper ability to participate.

On the flipside, RRI (and RRI-aligned approaches) can provide significant advantages to organisations looking to innovate responsibly and demonstrate that responsible behaviour to their stakeholders (British Standards Institution, 2020). Similarly, focus on RRI in a Defence-setting does not just help secure safe, reliable, and effective technologies but also ensures the developers, Commanders, and operators of those technologies identify, acknowledge, and work to mitigate ethical risks. For one example that uses RRI to forecast potential ethical and policy issues in relation to security- and Defence-focused technologies, see Inglesant, Jirotka, and Hartswood (2018).

The context of using autonomous systems in Australian Defence adds additional layers of governance. These are chiefly informed by Australia's commitment to international humanitarian law (IHL), which requires a range of measures to reduce the adverse humanitarian effects of warfare. When it comes to adopting new technology, Article 36 of Additional Protocol I to the Geneva Convention creates a specific anticipatory obligation:

> *In the study, development, acquisition or adoption of a new weapon, means or method of warfare, a High Contracting Party is under an obligation to determine whether its employment would, in some or all circumstances, be prohibited by this Protocol or by any other rule of international law applicable to the High Contracting Party.*

Article 36 reviews typically occur late in the acquisition process for Defence, which does not allow for an iterative design process amongst stakeholders. An alternate methodology for Article 36 reviews, which begins in early design and development stages, provides an opportunity ensure better legal compliance and ultimately to secure more humanitarian and ethical outcomes. As we describe below, this method supports Article 36 reviews by establishing a process for dialogue between Defence and Defence industry during design and acquisition phases of systems.

Here, we use three key points from core RRI literature to inform our discussion. First, that responsibility requires embedding ethical principles within the design of technology and applying the precautionary principle (European Environment Agency, 2001; von Schomberg, 2013). Second, that the four dimensions for RRI/RI (anticipation, reflexivity, inclusion, and responsiveness) provide a means for making explicit modes of innovation governance (Stilgoe, Owen, & Macnaghten, 2013). And, third, that operationalising RRI methods relies on working closing with key stakeholders and understanding their conceptions of responsibility (Glerup, Davies, & Horst, 2017) and accountability (Srinivasan and González, 2022) to prevent disengagement and enhance empathy.

In focusing on an applied industry-led example, we also address an ongoing critique of the RRI which is that "most research [projects in RRI] … have been about RR [responsible research] and not RI [responsible innovation]" (Tait, 2017). This critique centres on the lack of engagement with downstream innovation and calls for a deeper understanding of innovation processes in technology sectors. Our case study helps answer this call by presenting an industry-led approach outside of academic innovation and focuses on an example with practical, downstream outcomes.

In the remainder of this article, we introduce the autonomous system and the project team, followed by an overview of the approach taken for this case study. We discuss the outcomes of this approach and lessons learnt for the next stages of our work.

### *Introducing Athena AI*

Athena AI is a targeting evaluation frameset – a tool to reduce cognitive workload for operators and improve compliance with the Defence-centric legal and ethical frameworks. It works by identifying objects and people who cannot be targeted on the battlefield. The intended result of using Athena AI is improving outcomes from a humanitarian perspective. Athena AI is produced by an Australian-based small-to-medium enterprise (SME) who worked with Trusted Autonomous Systems to design the system's ethical and legal support system. The industry team who developed this capability include ex-military personnel, leading technologists, and a data scientist. They worked collaboratively and iteratively with scientists, philosophers, military ethicists and legal researchers from Defence Science and Technology Group (DSTG) and universities to produce a novel solution for complex military decision making.

The technology is designed to address the highly dynamic nature of targeting in military operations. A good example is targeting a moving vehicle as it goes through a built up urban environment. Whoever decides on an attack on the vehicle must consider the impact point of the weapon used, the primary effects of the weapon on the vehicle, and incidental damage to nearby people, vehicles, buildings, and infrastructure. This assessment changes constantly depending on the movement of the targeted vehicle and its proximity to protected persons and objects.

### *Approach: A Method for Ethical AI in Defence*

The ethical foundations of a minimally just application of artificial intelligence in autonomous weapons systems (Galliott and Scholz, 2018) and ethics-by-design principles (Shilton, 2013; Dignum, 2018; d'Aquin, Troullinou, O'Connor, Cullen, Faller, & Holden, 2018) were embedded from the start of the project in 2018. 'Improved ethical decision making' was identified as a core technical performance indicator and evaluable measure of success. Ethical and legal design processes were tried and applied through several workshops during the first twelve months of the project in 2019. Some of the ethical methods of the project translated into pragmatic tools for ethical AI, featured in a report published by Australian Defence in 2021, 'A Method for Ethical AI in Defence' (MEIAD) (Devitt, Gan, Scholz, & Bolia, 2021).

MEAID incorporates evidence-based hypotheses represented as topics and five facets of ethical AI in Defence drawn from over 100 attendees of a workshop from 45 organisations. It also presents pragmatic approaches for delivering on Australian commitments to develop, deploy, and govern trustworthy AI informed by global best practise, input and feedback from workshop participants and lessons from this case study. The report compares findings of the workshop to existing ethical AI frameworks including Australia's national AI ethics principles and commitment to Article 36 reviews of all new means and methods of warfare (Commonwealth of Australia, 2018).

MEIAD presents five facets for creating ethical AI. The five facets are: responsibility (who is responsible for AI?), governance (how is AI controlled?), trust (how can AI be trusted), law (how can AI be used lawfully?), and traceability (how are the actions of AI recorded?). The report also features tools that align with risk documentation requirements for systems developers. The addition of MEAID aims to ensure that developers identify, acknowledge, and work to mitigate ethical risks during design processes and the testing and evaluation of AI.

As outlined below in 'Applying the Method for Ethical AI', the adoption of the approaches outlined in the report produced activities closely align with four dimensions for responsible innovation identified by Stilgoe et al. (2013). An example of links between dimensions and activities is presented in Table 1.

*Table 1: Athena AI case study and the four dimensions of responsible innovation*

| Dimension of responsible innovation (Stilgoe et al., 2013) | Indicative techniques and approaches (Stilgoe et al., 2013) | Approach in Athena AI case study |
|---|---|---|
| Anticipation | Foresight; scenarios; vision assessment | Scenario development and validation by external experts |
| Reflexivity | Multidisciplinary collaboration and training; Embedded social scientists and ethicists; Codes of conduct | Multidisciplinary, multisector collaboration during legal and ethical workshops |
| Inclusion | Focus groups, deliberative mapping; user-centred design | User-centred design aimed to authentically represent the problem space, reduce cognitive load and support ethical decision-making |
| Responsiveness | Regulation; Standards; Stage-gates; Value-sensitive design | Abidance by ethical and legal frameworks as described in approach; Responsive to the values and expectations of end users and stakeholders. Use of ethical risk matrix at project milestones and ongoing throughout |

***Applying the Method for Ethical AI***

The system at the centre of this Athena AI case study is an object classifier which identifies signs of surrender, non-combatants, protected symbols, and signs of injury. These identifiers are used to alert an operator that an object or person cannot be targeted on the battlefield. The AI classifier and software development tool combines a legal and ethical framework and presents an AI decision support tool that enhances compliance with IHL.

**Scenario development:** The initial step for developing Athena AI was responding to scenarios provided by a former military legal officer from the Australian Defence Force and reviewed by academic researchers, Defence scientists, and industry. These scenarios presented examples of difficult decisions made on limited intelligence in addition to elements that must be considered under IHL. These scenarios helped the project team establish the initial framework for the user interface and establish the rules which would need to be implemented on the back end of the software.

**Legal and ethical workshops:** From there, the project team brought together a variety of moral philosophers, ethicists, lawyers, and human factors experts. Together, the team and experts considered how AI might support the implementation of IHL and rules of engagement, then considered what this might look like in code. These discussions produced a 70-page legal and ethical framework and a plan for how to integrate the framework with Athena AI. This framework is reviewed every six months based on AI neural network performance, any changes in law and input from weapons review experts.

**Data Ethics Canvas:** The Data Ethics Canvas by the Open Data Institute[1] was an initial tool that allowed engineering teams, non-engineering teams, and military personnel to consider a variety of ethical aspects of data. This included: data collection, ways for reviewing and preventing bias, and means for tracing data back to its original source for machine learning models. The Data Ethics Canvas helped the team consider the end-user and who might be associated with the results of the AI. This included questions that included: What happens if something goes wrong? What if there is biased data? What happens if the system mislabels something? How might these be mitigated?

**Ethical AI risk matrix:** The ethical risks flagged in the Data Ethics Canvas contributed to the project team's ongoing ethical AI risk matrix. This matrix tracks ethical risks, flags actions needed to mitigate those risks, and identifies who is responsible for those actions and when that mitigation will occur. At every milestone of the project and periodically, the team reviews these ethical risks to understand whether they have been mitigated, closed out due to design choices, or whether they are ongoing and require monitoring.

### *Outcomes: Establishing trust and improving decision making*
Understanding and responding to ethical and legal considerations was essential for the development of Athena AI. In the table below (Table 2) we present the initial outcomes for the project in relation to the five facets of a Method for Ethical AI and dimensions of RRI/RI (Stilgoe et al., 2013). These outcomes informed the ethical foundations for the design team and system requirements; providing a framework and roadmap for iterative design and development.

The team is building on these findings as the Athena AI system through test, evaluation, verification and validation (TEVV) towards future development. Some of this testing includes independent human factors evaluation of the performance of Athena AI by the Human-Autonomy Research Team at Defence Science & Technology Group and research by the Centre for Human Factors and Sociotechnical Systems, University of the Sunshine Coast, to map the Australia's Military Military Ethics Doctrine[2] and MEAID to create best practice evaluation metrics.

---

[1] The Date Ethics Canvas, https://theodi.org/article/the-data-ethics-canvas-2021/

[2] ADF-P-0 Military Ethics https://theforge.defence.gov.au/adf-philosophical-doctrine-military-ethics

*Table 2: Outcomes of a Method for Ethical AI in Defence for Athena AI*

| MEAID facets | Outcome | Dimensions of RRI/RI |
|---|---|---|
| Responsibility | The AI has been implemented so that humans bear ethical and legal responsibility for decisions and the AI helps provide clear, actionable information to the decision makers.<br><br>**Education for decision-makers and operators:** Responsibility requires education and training so decision-makers and operators understand when to trust and when to cross-check information provided to them by the AI as well as awareness of how to interpret confidence ratings when the AI categorises persons and objects. | Reflexivity (training for Commanders and operators) |
| Governance | **Establishing a framework for governing the system:** By incorporating ethical considerations from early stages, the project team created a framework that could both govern the ongoing development of the AI and meet the legal and ethical requirements of a weapons review and Australia's systems of control (Commonwealth of Australia, 2019). | Responsiveness (establishment of framework to centre ethics-by-design) |
| Trust | **Creating a baseline for trust:** The design choices made to reduce ethical risk include additions in the training of the AI. These changes are vital for establishing a level of trust in the AI, which helps operators and decision-makers understand what level of trust they should have in the information it provides. Athena AI developed a full 'trusted' version control and validation framework which balances data into train, verify, and validate groups (Bradley, 2021). This ensures AI models are trained and tested on real world scenarios and done so with minimal bias. | Inclusion (user-centric design of the system) |
| Law | **Meeting legal requirements:** From a legal standpoint, the legal and ethical aspects of Athena AI are crucial for ensuring the system enhances compliance with IHL and passes an Article 36 weapons review. To address legal and ethical risks, Athena AI have worked extensively with International Weapons Review (Athena AI, 2021) to create the assurance framework and compliance with ADF's targeting cycle in order to be deemed lawful. | Responsiveness (abidance by international and national requirements) |

| Traceability | **Ensuring transparency:** A challenge posed by this process was the need to create new plans for identifying, tracking, and addressing ethical and legal aspects of the system. | Anticipation (use of scenarios to envision performance of the system) |
|---|---|---|
| | The records from this evolving discussion and design process have shaped future features of the system and opened the door to a wider range of applications beyond the original scope of the system. | |
| | Every neural network can be traced back to the data it was trained on, the parameters and metrics used for training, the performance results on unseen validation footage and any requirements for the network to pass. Additionally all decision logic is visible to the user such as blast damage probability, human frontal areas, and AI confidence levels. | |

*Operationalising RRI for AI in Defence*

Building ethical and legal autonomous systems in Defence is an iterative process – one which requires constant work and revision. Engaging in this process with care and responsiveness to stakeholder perspectives produced a tool that centres ethical and legal concerns and works towards the best possible humanitarian outcome. The design and development of this technology is an ongoing process, one that will continue to be shaped and challenged by the MEAID framework to ensure ethical risks are identified and addressed.

In this article, we have focused on the use of applied governance frameworks in the development of trustworthy autonomous systems. The success of this example of 'doing' RRI begins from the project team's acceptance that perfectionism is the enemy of progress. So far in the Athena AI project, we have successfully generated a genuine ethics-focused culture and encoded ethical and legal principles within the the system. Going forward, testing and evaluation will provide a robust review of this design and identify opportunities for improvement.

Returning to the three key guiding concepts adopted from RRI literature, we suggest, first, that this case study presents an example of how to mobilise responsibility by embedding ethical principles and anticipatory practices (technology foresight) within the design of Athena AI. Second, the processes used by the project team could be considered to have operationalised the anticipation, reflexivity, inclusion, and responsiveness dimensions of responsible innovation (see Table 1, above). This alignment is particularly strong in the case of the anticipation, reflexivity, and responsiveness dimensions. Finally, we suggest that this case study presents a clear example of the power of stakeholder engagement. The process of honestly and transparently engaging with stakeholders during different stages of this journey was essential to the outcomes achieved. The relationships built over this time were contingent on respectful engagement and transparency, commitment to action plans, and an ability to demonstrate attendance to issues as they arose. The reputation built through this process for Athena AI has led to business growth. It has also underscored the importance of charting flexible and practical paths for establishing trust in autonomous systems.

Considering the larger landscape of RRI, we suggest that operationalising "responsible innovation" rather than "responsible research" (Tait, 2017) in industry and Defence will require mandates for ethical design and assurance of technology, for instance in acquisition and testing and evaluation processes. The project team involved in this case study has established a community of practice for legal and ethical assurance of AI in Australian Defence to encourage such change. At the same time, we continue to build trust and ongoing relationships with industry to ensure we have a deep understanding of their innovation practices and to demonstrate how embedding RRI-aligned pricniples can have positive impacts for their operations.

**Acknowledgements:** Research for this paper received funding from the Australian Government through the Defence Cooperative Research Centre for Trusted Autonomous Systems.